# Multitone PSK Modulation Design for Simultaneous Wireless Information and Power Transfer

Prerna Dhull, *Graduate Student Member, IEEE,* Dominique Schreurs, *Fellow, IEEE,*
Giacomo Paolini, *Member, IEEE,* Alessandra Costanzo, *Fellow, IEEE,* Mehran Abolhasan, *Senior Member, IEEE,*
and Negin Shariati, *Member, IEEE*

*Abstract*—Far-field wireless power transfer, based on radio frequency (RF) waves, came into the picture to fulfill the power need of large Internet of Things (IoT) networks, the backbone of the 5G and beyond era. However, RF communication signals carry both information as well as energy. Therefore, recently, simultaneous wireless information and power transfer (SWIPT) has attracted much attention in order to wirelessly charge these IoT devices. In this paper, we propose a novel $N$-tone multitone phase shift keying (PSK) modulation scheme, taking advantage of the non-linearity of integrated receiver rectifier architecture. The main advantage of the proposed modulation scheme is the reduction in ripple voltage, introduced by the symbol transmission through phases. Achievable power conversion efficiency (PCE) and bit error rate (BER) at the output are considered to measure the efficacy of the proposed modulation scheme. Simulation results are verified by the measurements over the designed rectifier circuitry. The effect of symbol phase range, modulation order, and the number of tones are analyzed. In the future, this transmission scheme can be utilized to satisfy the data and power requirements of low-power Internet of Things sensor networks.

## I. INTRODUCTION

**R**ADIO frequency (RF) signals carry information as well as energy. Simultaneous wireless information and power transfer (SWIPT), which is a combination of wireless information transfer (WIT) and wireless power transfer (WPT), has recently gained prominence in making small Internet-of-Things (IoT) devices battery independent [1]. However, till now, most of the SWIPT research treats WIT and WPT performances separately, and either only WIT performance or only WPT performance is improved [2].

Recently, high peak to average power ratio (PAPR) signals have been shown to increase power conversion efficiency (PCE) in low-power WPT systems by exploiting non-linear characteristics of the rectifying circuitry [3]. Multitone waveform is one of the such high PAPR waveforms [4]–[6]. Several parameters such as tone separations, number of tones, input power range, bandwidth (BW), output filter cut-off frequency, output load, etc., have been shown to affect the achievable PCE at the output [7]–[15].

From an architecture perspective, most of the SWIPT research mainly revolves around a separated information and energy receiver architecture where the main focus is upon the distribution of the received power between information and energy paths according to a power utilization ratio, to optimize the SWIPT receiver performance [2]. Several techniques such as time-switching, power splitting, and frequency-splitting have been introduced to use the received signal in separate information and energy paths [16], [17]. Recently, a symbol-splitting technique has also been introduced for splitting the information and the carrier having no information, with the help of a coupler's isolation port to divide the received signal between information and power components in a better way for power transfer [18]. However, in such an architecture, the information is transmitted using conventional modulation techniques and the information detection is performed using conventional methods involving a local oscillator through a separate information path which involves increased power consumption for information detection at the receiver.

Therefore, an integrated information and energy receiver architecture as shown in Fig. 1(a) has been proposed where the same rectified output signal is used for information decoding as well as power extraction [19]. In this work, we leverage on this integrated information and energy receiver architecture. In fact, its main advantage is the reduction of the overall energy consumption at the receiver compared to the separated information and energy receiver architecture, by the removal of the RF local oscillator needed for information detection. However, in this case, conventional information transmission techniques cannot be utilized. Therefore, new transmission methods are required to be proposed for integrated information and energy receiver architectures.

To address this, a simple energy modulation scheme of transmitting symbols through different energy levels of a single-tone waveform for integrated receiver architecture is introduced [19]. To further increase the available constellation range for the energy symbols, an integrated receiver architecture model consisting of two half-wave rectifiers instead of having only one rectifier is proposed [20]. Two half-wave rectifiers are utilized with two amplitudes at the output, one posi-

P. Dhull is with RF and Communication Technologies (RFCT) Research Laboratory, University of Technology Sydney, Australia and Div. ESAT-WaveCoRE, KU Leuven, Belgium (e-mail: Prerna.Dhull@student.uts.edu.au).

D. Schreurs is with the Div. ESAT-WaveCoRE, KU Leuven, Belgium (e-mail: dominique.schreurs@kuleuven.be).

G. Paolini and A. Costanzo are with the Department of Electrical, Electronic and Information Engineering "Guglielmo Marconi" (DEI), University of Bologna, Italy (e-mail: giacomo.paolini4@unibo.it; alessandra.costanzo@unibo.it).

M. Abolhasan and N. Shariati are with RF and Communication Technologies (RFCT) Research Laboratory, University of Technology Sydney, Australia (e-mail: mehran.abolhasan@uts.edu.au; negin.shariati@uts.edu.au).

tive and one negative. Further, information is encoded in PAPR levels of multitone signals to enhance the WPT performance in [21], and its extended work demonstrated that non-uniform spaced multitone signals with fixed BW perform better in case of low SNR transmission scenarios compared to uniformly spaced multitone with varying BW [22]. Although [19]–[22] introduced new SWIPT transmission techniques for the integrated receiver architecture, their performance has not been verified through measurements. Another way of encoding the information is the use of pulse position modulation which also helps in increasing PAPR to boost harvested power [23].

Non-linearity of the rectifier causes intermodulations among multitone frequency components, resulting into the presence of ripples in the output voltage. A biased amplitude-shift-keying (ASK) waveform based upon these ripples optimized the SWIPT performance for integrated receiver architecture where each symbol carried some minimum energy for continuous energy harvesting at the receiver [24]. Further, a way of embedding information in amplitude ratios of individual tones of a multitone signal is introduced to make this amplitude-based information transfer immune to the transmission distance [25]. Another multitone amplitude transmission scheme has been introduced, utilizing two-dimensional signalling in terms of subcarrier amplitudes and the number of subcarriers [26]. The information is decoded from two separate paths with the help of current intensity and PAPR level. However, for such amplitude-based modulations, there should be a sufficient distinction in ripple voltage to identify the information symbols. Therefore, PCE varies significantly depending upon the transmitted symbols resulting into a significant dependence of WPT upon WIT transmission and hence, limiting the number of tones and modulation order for proper SWIPT performance.

Amplitude variations offer an advantage for WIT, but in integrated receiver SWIPT systems, it is required to minimize the effect of WIT over WPT. To minimize these ripples, a multitone frequency shift keying (FSK) scheme is introduced, where the tones' frequency spacings are varied according to the transmitted information, and information decoding is performed by Fast Fourier Transform (FFT) [27]. Another way of decoding information from multitone FSK is considering the fact that multitones with different frequency spacings result in different PAPR levels, which further helps in reducing the receiver power consumption by the removal of FFT [28]. A way of utilizing the phase of 2-tone signal is introduced in [25]. However, in this work, the method of information decoding is performed by analysing the DC and AC components at the output. This way of information transmission and reception becomes very complex even with a signal of three tones. Therefore, the proposed method in [25] is not suitable for an Orthogonal-Frequency-Division-Multiplexing (OFDM)-type communication where multiple symbols can be transmitted over the same signal. Also, it does not analyze its WIT performance as the focus is solely on WPT performance. A comparison among the existing transmission schemes for separated information and energy receiver architecture and integrated information and energy receiver architecture has been performed in Table 3 and Table 4 of [1], respectively.

In this paper, we introduce a novel multitone phase shift keying (PSK) modulation scheme for the integrated information and energy receiver architecture. Power, as well as information detection, is performed over the rectified signal. Information is transmitted in the phase differences between consecutive tones of multitone RF signal, and the information decoding at the receiver is performed by observing the phases of corresponding baseband tones of the rectified signal. Figure-of merits considered for analysing WPT and WIT performances are PCE and bit-error-rate (BER), respectively. The effect of the chosen symbols constellation, modulation order, and the number of tones over signal PAPR, PCE, and BER are observed, showing a requirement for an optimum design of symbol phases. Further, the simulation results are verified by measurements with a designed receiver rectifier circuitry.

The main advantage of proposed multitone PSK scheme is the transmission of $(N-1)$ symbols over an $N-$tone multitone signal which can be closely related to the standard OFDM used for WIT. This entire SWIPT transmission is performed using a simple receiver rectifier circuitry having low power consumption, as information detection does not require an RF local oscillator. Furthermore, the effect of WIT on the WPT performance of a SWIPT system has been reduced with the help of information transmission in form of tones' phases instead of amplitudes of the multitone signal. In addition, multitone PSK signal improves the end-to-end SWIPT performance by reducing the chances of saturation of the transmitter power amplifier. Table. I in Section V compares the proposed multitone PSK with the existing transmission schemes for integrated information and energy receiver architecture.

This paper is organized as follows. Section II introduces the theoretical model of multitone PSK signal design with non-uniform spaced tones. Section III illustrates the relationship between the chosen phase range for symbol constellation and the multitone PSK signal PAPR. Next, in Section IV, we present the rectifier design and measurement setup used for simulations and measurements. Then, the modulation scheme impact on WIT and WPT performances of the SWIPT system is analysed in Section V. Further, the obtained results are verified with the measurements. In the end, a conclusion is drawn in Section VI.

## II. Signal Model

The system model is illustrated in Fig. 1 where multitone PSK RF signal $x(t)$ is passed through the rectifier circuitry consisting of an input matching network, diodes, and an $RC$-LPF for both WIT and WPT. $x(t)$ consists of $N$ frequency tones at the centre frequency $f_c$ of 2.45 GHz. After passing through the rectifier, $x(t)$ results in the baseband output signal $y(t)$. $y(t)$ is used for both power transfer as well as information decoding. Complete $y(t)$ is used for WPT while information is decoded from phases of some of the baseband tones as shown in Fig. 1(b).

The modulation scheme is designed by making use of the non-linearity of the rectifier. The idea of the multitone PSK comes from the fact that an $N$-tones multitone signal passing through the rectifier results in intermodulation frequency components of various orders with the dominant $2^{nd}$ order intermodulation (IM$_2$) component at the rectifier output.

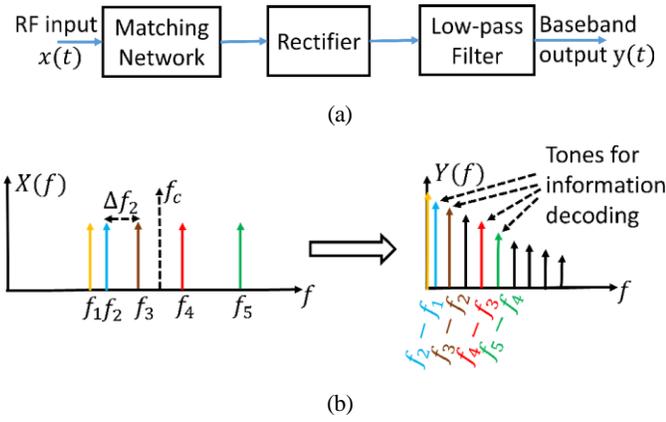

Fig. 1. (a) SWIPT integrated receiver rectifier and (b) 5-tone multitone PSK RF input signal spectrum $X(f)$ centred around carrier frequency $f_c = 2.45$ GHz and rectified baseband output spectrum $Y(f)$ consisting of IM frequency tones (in MHz) used for information decoding (coloured) and extra IM components (black).

IM$_2$s are generated by the mixing between two tones of $N$-tones multitone. In this paper, information is transmitted as phases of IM$_2$s between consecutive tones. The transmitted signal $x(t)$ can be considered as

$$x(t) = \text{Re}\left\{A \sum_{n=1}^{N} e^{j(2\pi f_n t + \phi_n)}\right\}, \quad (1)$$

where $A$ and $\phi_n$ represent the amplitude and phase of the $n^{th}$ tone frequency $f_n$, respectively. In order to see the intermodulation behaviour, a case of 3-tones with $N = 3$ in (1) can be considered. After transmitting the 3-tone multitone through the model of Fig. 1(a), the filtered output signal $y(t)$ consists of dc and a combination of baseband intermodulation frequency components. All the harmonics and odd order intermodulation components are filtered out by the LPF as all these frequencies would be at RF frequencies. Among baseband tones, IM$_2$ would be dominating compared to $4^{th}$ order, $6^{th}$ order, $\cdots$, etc., frequency components [29]. Therefore, $y(t)$ can be represented in terms of dc and IM$_2$s as

$$y(t) = \text{dc} + A_1 \cos(2\pi(f_2 - f_1)t + \phi_2 - \phi_1)$$
$$+ A_2 \cos(2\pi(f_3 - f_2)t + \phi_3 - \phi_2)$$
$$+ A_3 \cos(2\pi(f_3 - f_1)t + \phi_3 - \phi_1), \quad (2)$$

where $A_1$, $A_2$ and $A_3$ represent the amplitudes of IM$_2$ at $(f_2 - f_1)$, $(f_3 - f_2)$, and $(f_3 - f_1)$, respectively. In this paper, the signal is designed in such a way that the information is in the form of phases of IM$_2$ between consecutive tones only, i.e., $(f_2 - f_1)$, $(f_3 - f_2)$, $\cdots$, $(f_N - f_{N-1})$ tones phases would carry the information. In this way, we are able to transmit $(N - 1)$ information symbols over $N$-tone multicarrier signal as shown in Fig. 1(b). The 5-tone multitone PSK RF signal $x(t)$ results in a baseband signal consisting of various intermodulation frequency components of which four IM$_2$ frequency components generated by intermodulations between consecutive frequencies are used for phases information detection. However, multitone frequency spacings ($\Delta f_n$) and $\phi_n$ are required to be properly designed to make this multitone PSK signal feasible. Thus, the required constraints related to multitone PSK frequencies and phases are discussed in Sections II-A and II-B, respectively.

*A. Multitone PSK Frequencies*

In order to transmit $(N - 1)$ symbols with only a single transmission of multitone signal, $f_n$'s are chosen in such a way that $(N - 1)$ IM$_2$s between consecutive frequencies do not overlap each other and also do not coincide with other non-consecutive IM$_2$s. This results in unique $(N - 1)$ desired baseband frequencies at the output, and the information symbols are decoded by observing phases of these IM$_2$. Therefore, multitone $f_n$'s can not be uniformly spaced in (1), and require an appropriate selection of $\Delta f_n$.

The $n^{th}$ tone frequency of multitone PSK signal can be represented by

$$f_n = f_{n-1} + \Delta f_{n-1}, \qquad n = 2, 3, \cdots, N \quad (3)$$

where $\Delta f_{n-1}$ refers to frequency spacing between $n^{th}$ and $(n - 1)^{th}$ tone. $\Delta f_{n-1} \neq \Delta f_n \ \forall \ n = 2, 3, \cdots, N$ to distinguish the desired $(N - 1)$ IM$_2$s generated by consecutive frequencies of multitone PSK signal and must satisfy

$$\Delta f_{n-1} \notin \boldsymbol{U_i} \times \Delta \boldsymbol{F_i}, \quad \forall \ n = 2, 3, \cdots, N \text{ and}$$
$$i = 1, 2, \cdots, n - 1 \quad (4)$$

condition where $\boldsymbol{U_i}$ represents an upper triangular matrix of order $i \times i$ having non-zero elements as 1

$$\boldsymbol{U_i} = \begin{bmatrix} 1 & 1 & 1 & \cdots & 1 \\ 0 & 1 & 1 & \cdots & 1 \\ 0 & 0 & 1 & \cdots & 1 \\ \vdots & \vdots & \vdots & \ddots & \vdots \\ 0 & 0 & 0 & \cdots & 1 \end{bmatrix}, \quad (5)$$

and $\Delta \boldsymbol{F_i}$ is a column vector as

$$\Delta \boldsymbol{F_i} = \begin{bmatrix} \Delta f_1 \\ \Delta f_2 \\ \vdots \\ \Delta f_i \end{bmatrix}. \quad (6)$$

Here, multitone PSK is considered to be centred around centre frequency $f_c$, and $\Delta f_1$ is assumed to be the greatest common divisor ($GCD$) among $f_n$'s. Appropriate $f_n$'s can be determined by Algorithm 1. A BW spreading factor $r$ is used to have the signal with minimum BW ($r = 0$) and wider BW ($r > 0$) for a particular $N$. Hence, overall signal BW not only is a factor of $N$ but also depends upon the chosen values of $GCD$ and $r$. For example, for $N = 5$, $f_c = 2.45$ GHz, $GCD = 1$ MHz, and $r = 0$, $\Delta f_n$s would be 1 MHz, 2 MHz, 4 MHz, and 5 MHz from Algorithm 1. Therefore, the corresponding multitone PSK signal $x(t)$ would be having tones at 2.444 GHz, 2.445 GHz, 2.447 GHz, 2.451 GHz, and 2.456 GHz. After passing this multitone PSK signal through the rectifier, baseband signal having various IM frequency components spaced at minimum 1 MHz apart are generated as GCD = 1 MHz. Four IM frequency tones at 1 MHz, 2 MHz, 4 MHz, and 5 MHz are utilized for information detection.

$\Delta f_n$ increases with an increase in $GCD$ or $r$, which is helpful from WIT perspective. The reason for this is that it is relatively easier to identify the required wider-spaced baseband tones at the output. However, rectifier's matched BW and LPF's cut-off frequency ($f_{cutoff}$) imposes a limit over allowable multitone $N$, $GCD$, and $r$ for proper WPT.

**Algorithm 1** Multitone PSK Frequencies

**Input:** $f_c$, $N$, $GCD$, $r$
**Output:** $\Delta f_n$, BW, $f_n$
  *Initialisation* : $k_1 = 1$, $i = 1$, $c = 1$
1: $\Delta f_1 = k_1 * GCD$, $\Delta f_1 = GCD$
  *LOOP Process*
2: **for** $n = 2$ to $N - 1$ **do**
3:   $K$ = all combinations of $k_n = U_j.k_{j-1} \forall \quad j = \{1, 2, \cdots, n-1\}$
4:   **while** ($i \in K$) **do**
5:     $i = i + k_1 + r$
6:   **end while**
7:   $k_n = i; i = 1; \Delta f_n = k_n * GCD$
8: **end for**
9:   BW = $\sum_{n=1}^{N-1} \Delta f_n$
10: Symmetrically align frequencies around $f_c$
11: **return**

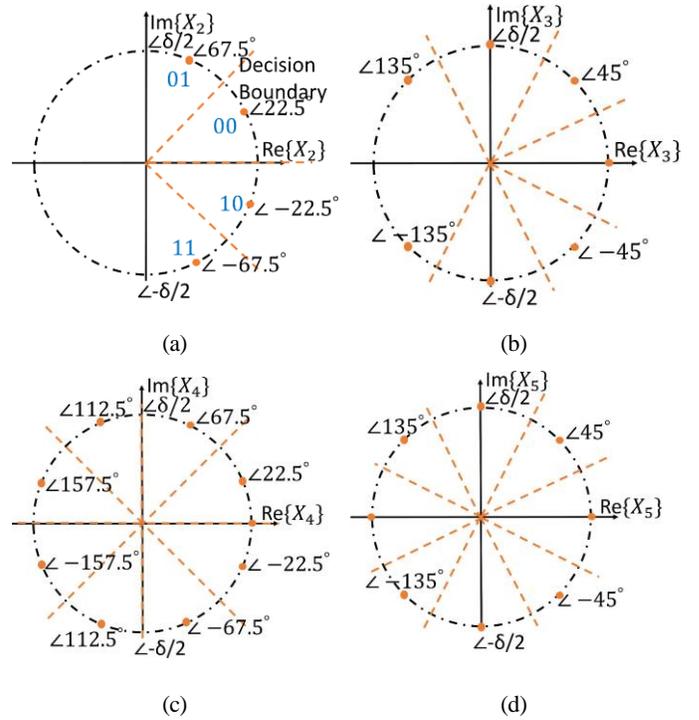

Fig. 2. Multitone PSK tone phase constellation for $n^{th}$ tone, (a) $n = 2$, (b) $n = 3$, (c) $n = 4$, and (d) $n = 5$, for $M = 4$ and $\delta = [-90°, 90°]$.

### B. Multitone PSK Phases

From (2), it can be observed that the output baseband tones phases are phase differences between the corresponding multitone frequencies. Therefore, in multitone PSK, symbols are encoded as the phase differences of the consecutive tones. Hence, $n^{th}$ tone phase $\phi_n$ can be represented in terms of transmitted information symbols as

$$\phi_n = \sum_{i=1}^{n-1} \phi_i + s_{n-1}, \qquad n = 2, 3 \cdots, N \qquad (7)$$

with the assumption of the first tone phase, $\phi_1 = 0$. In (7), $s_{n-1}$ represents the information symbol, transmitted as the difference between $\phi_n$ and $\phi_{n-1}$.

Let $\Phi_x$ and $\Phi_y$ be the lower and upper bound of symbols constellation such that $\Phi_y - \Phi_x = \delta$, where $\delta$ is defined as a considered phase range to allocate the information symbols within this range as shown in Fig. 2(a). Thus, the available information symbols set **S** containing $M$ symbols can be defined as

$$\mathbf{S} = \Phi_x + (2m-1)\frac{\delta}{2M}, \quad \forall \quad m = 1, 2, \cdots, M \quad, \quad (8)$$

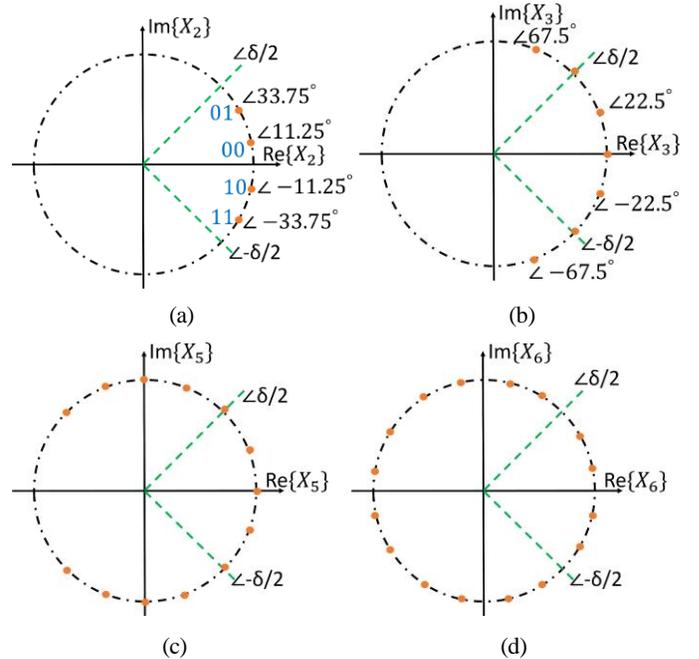

Fig. 3. Multitone PSK tone phase constellation for $n^{th}$ tone, (a) $n = 2$, (b) $n = 3$, (c) $n = 5$, and (d) $n = 6$, for $M = 4$ and $\delta = [-45°, 45°]$.

in terms of $\delta$ and modulation order $M$. Transmitted $(N - 1)$ information symbols over an $N$-tone multitone PSK signal would belong to this available information symbols set **S**.

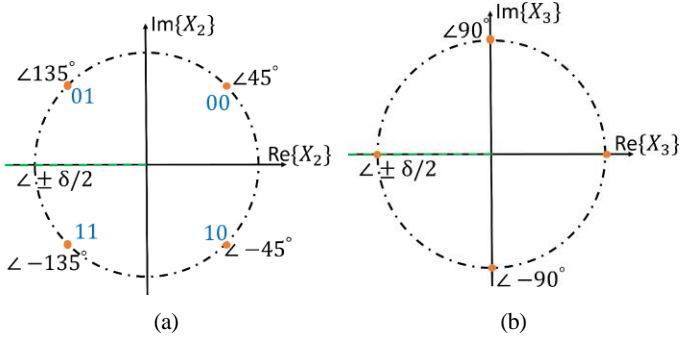

Fig. 4. Multitone PSK tone phase constellation for $n^{th}$ tone, (a) $n = 2$ and (b) $n = 3$, for $M = 4$ and $\delta = [-180°, 180°]$.

Assuming the symbols to be equidistant and symmetric to $x$-axis, i.e., $\Phi_x = -\delta/2$, (8) can be rewritten as

$$S = (2m - M - 1)\frac{\delta}{2M}, \quad \forall \ m = 1, 2, \cdots, M \quad . \quad (9)$$

Therefore, for $\delta = [-90°, 90°]$ and $M = 4$, available symbol phases set $S$ would be $67.5°$, $22.5°$, $-22.5°$, and $-67.5°$, as shown in Fig. 2(a). Each symbol carries $\log_2 M$ bits.

Actual possible individual tone phases $\phi_n$ are different from the chosen transmitted symbol from $S$, as each tone has different choices for its phase as seen in (7), varying according to $\delta$ and $M$. As $\phi_1$ is assumed to be zero, from (7), $f_2$ can have its phase $\phi_2$ as

$$\phi_2 \in S, \quad (10)$$

and depicted in Fig. 2(a) for $M = 4$ and $\delta = [-90°, 90°]$. Further, from (7), $\phi_3$ can be seen as a combination of two symbols as

$$\phi_3 \in \{s_i + s_j, \quad \forall \ i, j = 1, 2, \cdots, M\}, \quad (11)$$

where $s_i$ and $s_j$ are symbols from $M$ constellation points defined by (9). (11) can be rewritten as

$$\phi_3 \in (i + j - M - 1)\frac{\delta}{M}, \quad \forall \ i, j = 1, 2, \cdots, M \quad (12)$$

resulting in $(2M - 1)$ available phases for $\phi_3$ shown in Fig. 2(b), as $(i + j)$ varies between $[2, 2M]$. Generalizing, $f_n$ has $((n - 1)M - n - 2)$ phase combinations, varying between $-(n - 1)(M - 1)\frac{\delta}{2M}, (n - 1)(M - 1)\frac{\delta}{2M}$.

However, as seen in Fig. 2, these obtained phase combinations are not all different from each other as phases are overall restricted to lie within the interval of $[-180°, 180°]$. Therefore, phase combinations of (7) start repeating itself after a particular $n^{th}$ tone satisfying

$$(n - 1)(M - 1)\frac{\delta}{2M} \geq 180°, \quad (13)$$

and converge to maximum total available tone phases choices of $(360°/(\delta/M))$. Fig. 2, Fig. 3, and Fig. 4 illustrate this phase combining behaviour for $\delta$ as $[-90°, 90°]$, $[-45°, 45°]$, and $[-180°, 180°]$, respectively for $M = 4$. It can be observed that for a particular $M$, smaller the $\delta$, larger is the $n^{th}$ tone after which the tones would be able to have phases distributed in the whole $[-180°, 180°]$ interval. This comparison can be analysed by observing Fig. 2(c), Fig. 3(d), and Fig. 4(b) where phase constellation points start overlapping each other after $4^{th}$, $7^{th}$ and $3^{rd}$ tone for $\delta$ as $[-90°, 90°]$, $[-45°, 45°]$, and $[-180°, 180°]$, respectively, for $M = 4$.

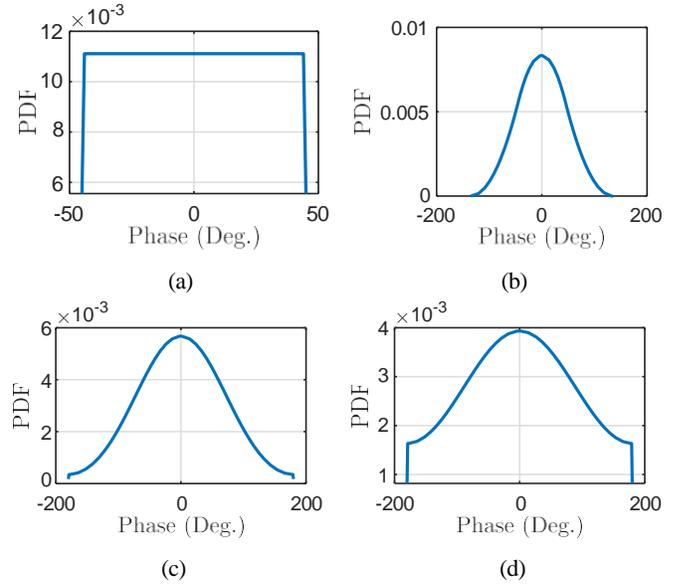

Fig. 5. Probability density function (PDF) for $n^{th}$ tone of multitone PSK for (a) $n = 2$, (b) $n = 4$, (c) $n = 8$, and (d) $n = 16$ for $\delta = [-45°, 45°]$.

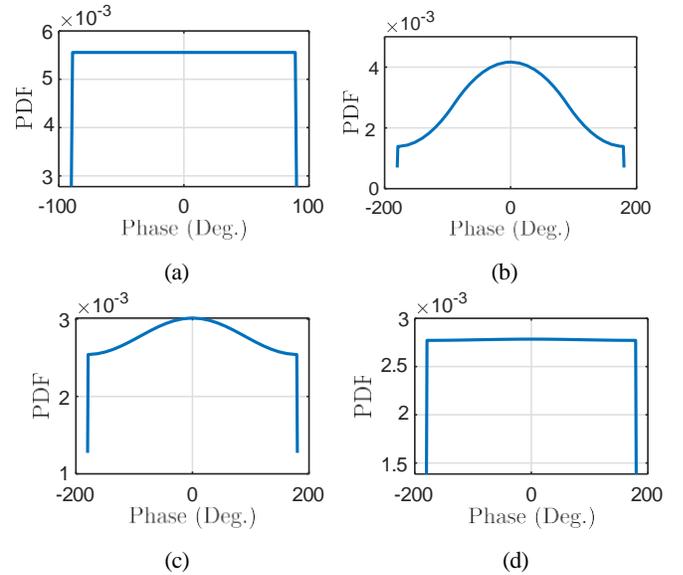

Fig. 6. Probability density function (PDF) for $n^{th}$ tone of multitone PSK for (a) $n = 2$, (b) $n = 4$, (c) $n = 8$, and (d) $n = 16$ for $\delta = [-90°, 90°]$.

## III. MULTITONE PSK PHASE DISTRIBUTION AND PAPR

In multitone PSK, as the information is being transmitted in the phases, it is important to analyse the effect of these introduced phases upon WIT and WPT performances of the SWIPT system. Although a multitone signal having all its phases aligned performs best for WPT [4], the introduction of phases provides the benefit of transmitting the information to realize a SWIPT operation. Therefore, we need to analyse the extent upto which $\phi_n$'s of multitone PSK affect WPT performance. Here, in this section, we study the effect of phases over the signal PAPR which is one of the figure of merits for WPT performance evaluation [30].

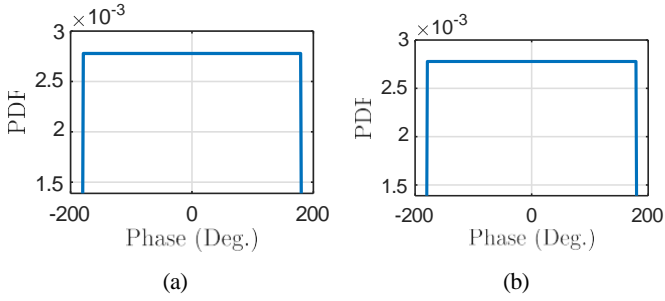

Fig. 7. Probability density function (PDF) for $n^{th}$ tone of multitone PSK for (a) $n = 2$ and (b) $n = 3$ for $\delta = [-180°, 180°]$.

In this paper, symbol constellations are assumed to be symmetric along x-axis as discussed in the Section II-B. In this way, multitone signal would have higher probability of having phases closer to zero which is beneficial for WPT, as described in further discussion.

Let $\mathbf{S}$ defined by (9) be uniformly distributed between $[-\delta/2, \delta/2]$, i.e., equiprobable symbols with mean 0 and variance $\delta^2/12$. From (7), phase of an $n^{th}$ tone can be seen as the sum of $(n - 1)$ symbol phases. Thus probability density function (PDF) of $\phi_n$ can be evaluated by convolving the $(n - 1)$ uniformly distributed PDF's. Therefore, PDF of $\phi_n$ follows an Irwin-Hall distribution ranging from $[-(n-1)\delta/2, (n-1)\delta/2]$ with mean 0 and variance $n \times \delta^2/12$ [31]. It is well known that Irwin-Hall distribution closely resembles the Gaussian distribution as $n$ increases [31], resulting in higher probabilities of occurrence of phases closer to zero, as seen in Fig. 5(b).

However, as discussed in previous section, constellation points for a $\phi_n$ starts repeating itself after a particular $n$ due to $[-180°, 180°]$ constraint. Therefore, all tone phases $\phi_n$ do not follow Gaussian distribution and the overall PDF can be found by applying

$$\Pr(x) = \sum_{k=-\infty}^{\infty} \Pr(2\pi k + x), \quad (14)$$

over obtained phases, where $\Pr(x)$ represents the probability at $x$.

Fig. 5 illustrates the PDF behaviour of $\phi_2$, $\phi_4$, $\phi_8$, and $\phi_{16}$ for $\delta = [-45°, 45°]$. It can be seen that PDF for $\phi_2$ and $\phi_4$ closely resembles Irwin-Hall distribution, whereas PDF for $n \geq 7$ starts getting modified as probability of having phases closer to 180° increases. As $n$ increases, more and more phases combinations would occur farther from zero phases moving towards uniform distribution PDF as shown in Fig. 5(c). The $n^{th}$ tone of multitone PSK after which PDF starts following uniform distribution depends upon the choice of $\delta$. From Fig. 6, we can see that PDF of $\phi_n$ starts moving towards uniform distribution at a faster rate for $\delta = [-90°, 90°]$ compared to $\delta = [-45°, 45°]$ (Fig. 5(d)), and a uniform distribution is achieved for $N = 16$. Further, Fig. 7 illustrates PDF for $\delta = [-180°, 180°]$ where all multitone PSK phases follow uniform PDF's. Therefore, larger the $\delta$, sooner the all $\phi_n$'s start following uniform distribution. In uniform distribution, the probability of having a larger phase (closer to 180°) is equal to the probability of having a 0° phase which would affect the transmitted signal PAPR and the obtained PCE at the output as discussed further.

PAPR for $x(t)$ can be obtained from

$$\text{PAPR} = \frac{\max\{|x(t)|^2\}}{\frac{1}{T}\int_{-T/2}^{T/2} x^2(t)dt}, \quad (15)$$

where $T$ is time-period of waveform $x(t)$. For an $N$-tone multitone signal, maximum PAPR of $2N$ can be attained by aligning all the phases [21]. However, phases are not aligned for the multitone PSK signal as information is needed to be passed in multitone phases. The PAPR for multitone PSK with varying $N$ is shown in Fig. 8. PAPR for each point, for example PAPR for $N = 6$, $M = 4$, $\delta = [-90°, 90°]$, is obtained by analysing 1000 transmitted multitone waveforms $x(t)$. Firstly, for each transmitted waveform $x(t)$, random binary information data are encoded according to the available symbol set $\mathbf{S}$ from (9). Then, $x(t)$ tones' phases are modified accordingly from (7). The input power level is set at $-10$ dBm for all the transmitted waveforms so that only the effect of tones' phases can be observed. Average of these 1000 obtained PAPR levels is considered to provide a PAPR value for a single case. Three cases of phase ranges $\delta$ are considered to analyse the phase distribution effect over the multitone signal PAPR with increasing number of tones.

It can be seen that multitone PSK input PAPR is lower compared to multitone having all $\phi_n = 0$, as expected. Further, as $N$ is increased for the multitone waveform, PAPR for a phase range gets saturated to maximum obtained PAPR. As $N$ for multitone PSK is increased, uniform distribution starts dominating for the most of the $\phi_n$, i.e., all phase combinations having equal probability. This means that now, even phases which are closer to 180° have same probability of occurrence as of phases closer to 0°. Therefore, overall PAPR reduces and saturates to a maximum upper value for a particular $M$ and $\delta$. It can also be seen that multitone PSK PAPR saturates earlier for larger $\delta$, i.e., $\delta = [-180°, 180°]$, resulting in the lowest PAPR, because $\phi_n$ starts following uniform distribution for all $n \geq 2$ for the case of maximum $\delta = [-180°, 180°]$.

Multitone signals are already being used for WPT transfer from quite some time. However, high PAPR signals deteriorates the overall end-to-end practical system performance as these signals may drive transmitter amplifier into saturation. However, PAPR of transmitted multitone PSK signal is not increasing linearly with $N$. Therefore, transmission of information in terms of tones' phases is offering an additional advantage of improving the overall SWIPT system performance.

### IV. RECEIVER MODEL

The simple rectifier model consisting of a matching network, diode circuitry followed by a LPF is designed. Further, the rectifier is fabricated and all multitone PSK simulation results are verified by the measurement results.

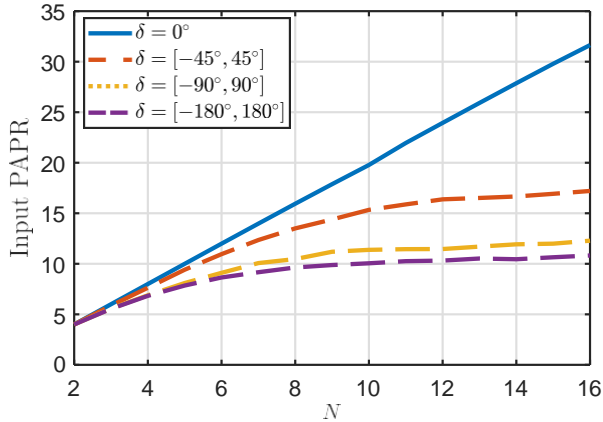

Fig. 8. Multitone PSK PAPR with varying $N$ for $M = 4$, $GCD = 1$ MHz and 1000 multitone streams for each $N$ and $\delta$.

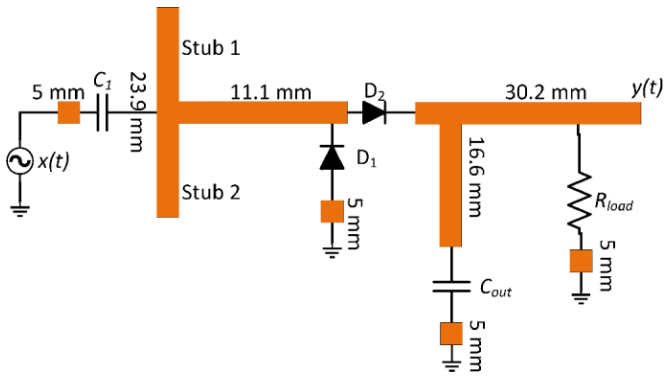

Fig. 9. Rectifier simulation model for SWIPT.

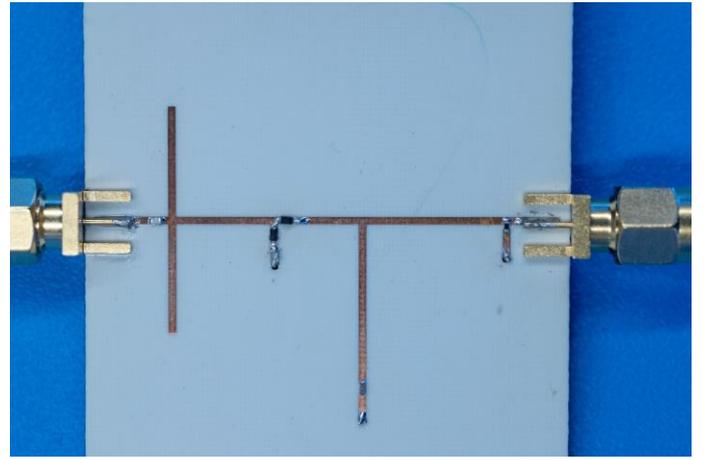

(a)

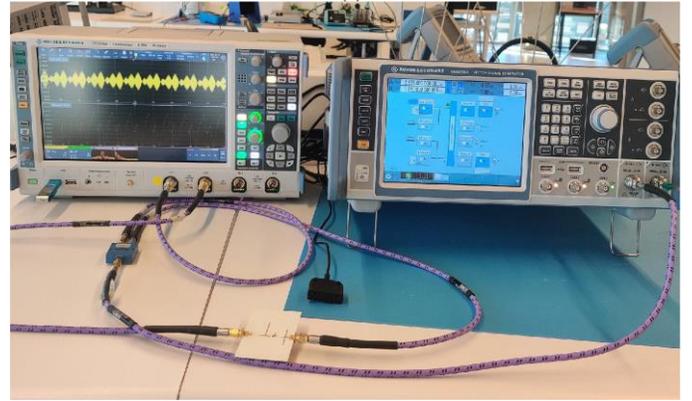

(b)

Fig. 10. (a) Fabricated integrated receiver rectifier and (b) Measurement setup with VSG R&S SMW200A, power splitter, rectifier, and oscilloscope R&S RTO2044.

### A. Rectifier design

A rectifier model depicted in Fig. 9 is designed solely to demonstrate the feasibility of the proposed multitone PSK for SWIPT operation and the topology is inspired from [32]. The topology consists of an input matching network with $C_1 =$ 1 pF, two Schottky diodes (Skyworks SMS7630-079LF), and a LPF ($C_{out} = 0.1$ pF and $R_{load} = 4.4$ kΩ). The rectifier has been designed to maximize PCE for comparatively larger matched BW of around 100 MHz centred on 2.45 GHz in terms of reflection coefficient ($S_{11}$) by the use of $C_1$, stub lengths, transmission lines lengths, and, $R_{load}$. This is required to make multitone PSK possible with larger $N$ to increase the throughput, i.e., for WIT performance of the SWIPT system. In short, the rectifier model has been designed keeping in mind WPT as well as WIT.

In the rectifier design, LPF cut-off frequency is selected in such a way that all the harmonics are filtered out by LPF, while simultaneously it should have sufficient BW to pass the relevant $IM_2$ baseband information tones. Therefore, the rectifier is designed to have a larger LPF BW than the highest relevant $IM_2$ frequency component, but smaller than the RF fundamental tones. For all the simulation and measurement results, $GCD$ and $N$ for multitone PSK signal are selected in such a way that the resulting signal would have a BW smaller than the RF BW.

The WIT and WPT performances of multitone PSK signal has been measured in terms of achievable PCE and BER, respectively, in order to analyze the overall SWIPT performance. PCE at the output can be defined as

$$\text{PCE} = \frac{|y_{dc}|^2/R_{load}}{P_{in}} * 100, \quad (16)$$

in terms of the transmitted power $P_{in}$ of $x(t)$ and dc voltage $y_{dc}$ of the received signal $y(t)$. PCE at the rectifier output varies according to the multitone PSK signal PAPR.

### B. Measurement Setup

The designed rectifier is fabricated on RO4350B substrate having dielectric constant as 3.66, loss tangent as 0.0031, and thickness of 0.762 mm and is shown in Fig. 10(a). The measurements are performed using the setup shown in Fig. 10(b). The transmitter and receiver are connected through the cabled connections. The modulated multitone PSK signals are generated using a vector signal generator (VSG) (R&S SMW200A). Further, the signal is divided into two streams using a power splitter. One stream is directly fed to the

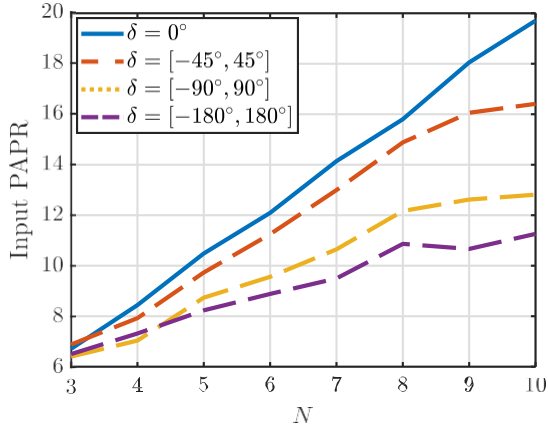

Fig. 11. Measured multitone PSK PAPR with varying $N$ for $M = 4$, $GCD = 1$ MHz, and 50 multitone streams for each $N$ and $\delta$.

Oscilloscope (R&S RTO2044 Channel 1) which is considered as the transmitted multitone signal reference from VSG as multitone PSK signal power has been reduced after passing through the power splitter. The other stream is passed through the rectifier and fed to the Oscilloscope Channel 2. Calibration for different cable lengths is performed before comparing the transmitted and the received signal from the oscilloscope for power and information analysis. As the information is in phases of frequency components, a spectrum analyser does not meet the requirement for information decoding. Therefore, an oscilloscope is used for the measurements. For the received signal, the input impedance of the oscilloscope has been set at a 1 MΩ value to minimize its effect over the rectifier's output impedance. This is performed to keep the measured performance similar to the simulated performance. Multitone PSK signals $x(t)$ with varying parameters $P_{in}$, $M$, $N$, $\delta$, and $GCD$ are generated and the demodulated signals $y(t)$ are analysed for the SWIPT performance in terms of attained PCE and BER. The effect of $\delta$ over the transmitted signal PAPR as discussed in Section III is verified by the measured PAPR over the Oscilloscope Channel 1 output and illustrated in Fig. 11. Multitone PSK signal streams are generated for each $N$ for each different allocated symbol phase range similar to the simulation analysis in Section III. It can be seen that the narrower the transmitted symbols' phase range, the better would be the WPT performance of the designed scheme as observed in simulated results in Fig. 8.

## V. Performance Analysis of Multitone PSK

To analyze the SWIPT performance of multitone PSK, PCE, and BER are used as figures of merit for WPT and WIT performances, respectively. Multitone stream frequencies and phases are designed as discussed in Section II-A and Section II-B. For example, for $N = 6$, GCD = 1 MHz, frequency spacings between consecutive tones are obtained as 1 MHz, 2 MHz, 4 MHz, 5 MHz, and 8 MHz from Algorithm 1. $x(t)$ would result into multitones having tones at 2.44 GHz, 2.441 GHz, 2.443 GHz, 2.447 GHz, 2.452 GHz, and 2.46 GHz. The time-domain transmitted waveform $x(t)$ and the corresponding received waveform $y(t)$ measured by the oscilloscope for the case of all tones' phases aligned and $P_{in} = -16$ dBm, are illustrated in Fig. 12(a).

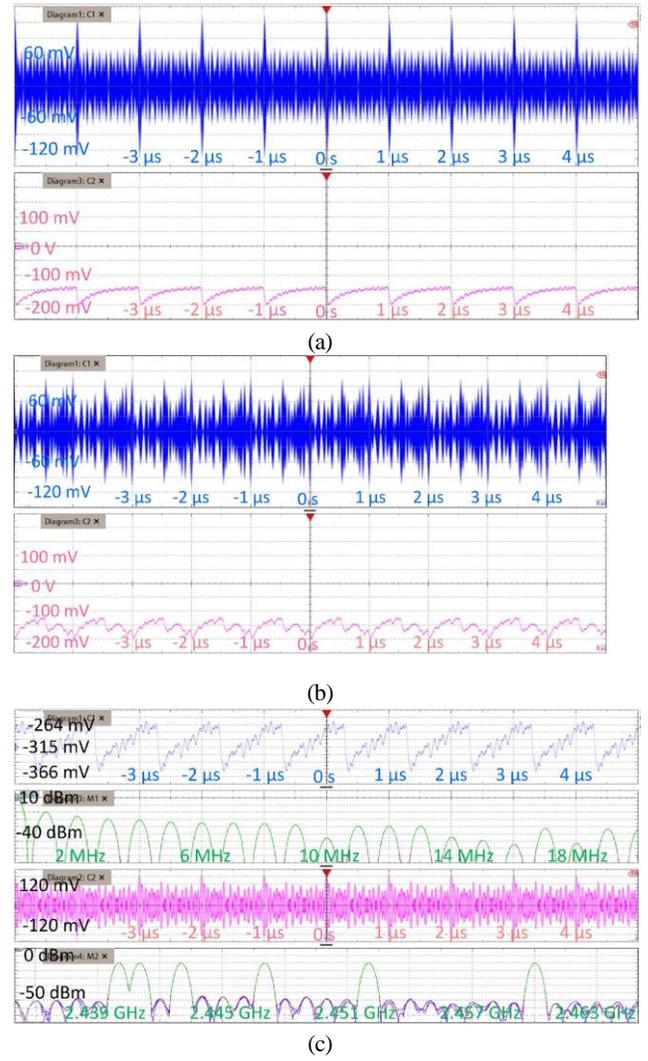

Fig. 12. Oscilloscope measured transmitted $x(t)$ and received $y(t)$ waveforms with (a) all phases aligned (b) with information symbol transmission, and (c) frequency spectra.

The phases of these tones would get modified according to the transmitted bit patterns in each multitone stream. For example, for $M = 4$ and $\delta = [-180°, 180°]$ in the above case with a transmitted bit stream of 1 0 0 1 0 1 0 1 0 0, the 6 tones phases would be 0°, − 45°, 90°, − 135°, 0°, and 45° by utilizing Fig. 4(a), (7), and (9). The symbols are encoded using the gray coding scheme to minimize the BER [31] as shown in Fig. 4(a). The corresponding transmitted waveform $x(t)$ and received waveform $y(t)$ measured by the oscilloscope are shown in Fig. 12(b). Another example of measured $x(t)$ and $y(t)$ with their respective frequency spectrum $X(f)$ and $Y(f)$ with random bit stream pattern, is illustrated in Fig. 12(c).

PCE and BER are analysed by obtaining the rectifier output time-domain waveform $y(t)$. Fig. 13 shows the obtained simulated PCE for a multitone PSK with $N = 6$, $M = 4$, $GCD = 1$ MHz, $r = 0$ for 100 transmitted multitone streams, i.e., 1000 bits. From Algorithm 1, BW of such signal is 20 MHz which is under the designed rectifier BW range. Hundreds of multitone streams for each scenario are transmitted where each time their phases are encoded according to a random information bit pattern. It can be seen that PCE is maximum for the aligned phases and decreases with increased $\delta$, as observed in the case of PAPR as well. Maximum attained PCE at $P_{in} = 0$ dBm, falls by around 3% when $\delta$ is increased from $0°$ to $[-180°, 180°]$, implying that simultaneous transmission of information in terms of multitone phases with the help of multitone PSK costs only around 3% of WPT performance when a WPT multitone signal is replaced by a multitone PSK signal for SWIPT transfer.

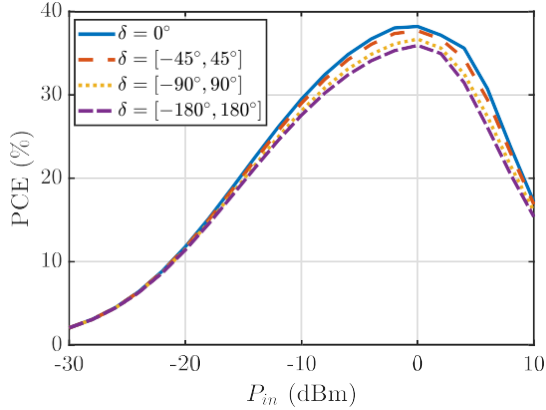

Fig. 13. Simulated PCE for multitone PSK centred around 2.45 GHz with $N = 5$, $GCD = 1$ MHz, and $M = 4$ for 100 multitone streams, i.e., 1000 bits.

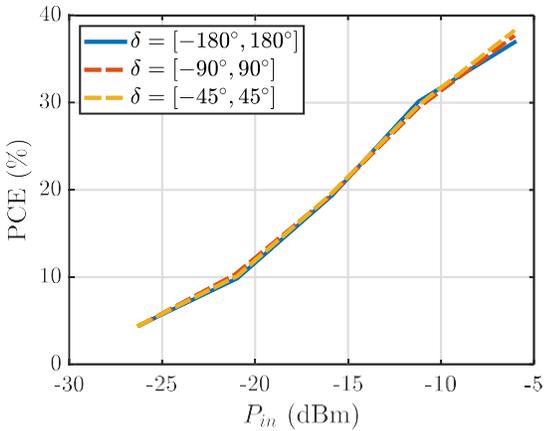

Fig. 14. Measured PCE for multitone PSK with $N = 6$, $GCD = 1$ MHz, and $M = 4$ for 100 multitone streams, i.e., 1000 bits.

Here, the peak PCE of the rectifier has been reduced somewhat to have a larger low-pass filter BW to simultaneously receive the DC signal for WPT as well as the baseband signal which contains the information, compared to conventional rectifiers that are utilized solely for WPT path (for separated information and energy receiver architecture). It is possible to design a rectifier to have 60% PCE but with only 0.5 MHz LPF BW [32]. On the other hand, a larger LPF BW is required to accommodate baseband tones carrying the information. Therefore, enabling the transmission of information is at the cost of some reduction in power conversion efficiency. Overall, it is well known that a trade-off exists between WPT performance and WIT performance in the case of integrated information and energy receiver architecture [19]. However, the overall system efficiency is increased as we are able to receive power while decoding information using the same rectifier, hence removing the need for additional downconversion electronics at the receiver. It is undoubtedly a low-cost solution, compared to others considered, without affecting the correct demodulation of the data.

The simulated results are further verified through measurements over the designed rectifier from Oscilloscope Channel 2 (measured $y(t)$). Fig. 14 illustrates PCE for $N = 6$, $M = 4$, $GCD = 1$ MHz, $r = 0$, and BW = 20 MHz for 100 transmitted multitone streams, i.e., 1000 bits stream over an input power range of around $-27$ dBm to $-6$ dBm. For the case of $-6$ dBm transmitted power, it is possible to receive $-10$ dBm power which results into PCE of 38%. Less variation in PCE in accordance with $\delta$ is an advantage of the proposed multitone PSK scheme. In the earlier amplitude-based modulation schemes, information transfer through amplitudes results into varying ripple voltage at the output with the varying symbol patterns. As the information is in phases for multitone PSK, only the tones' phases change for different transmitted bit stream patterns, while keeping their amplitudes constant. Therefore, output voltage does not change much with the information transfer which is a beneficial point from WPT perspective making power transfer performance less dependent upon information transfer for SWIPT systems.

To analyse BER of multitone PSK, the time-domain output waveform $y(t)$ of the rectifier is further processed in MATLAB in frequency-domain, to extract the phase information from the relevant baseband tones. For BER calculation of one scenario, hundreds of such multitone streams are transmitted where each time their phases are encoded according to a random information bit pattern as discussed above. From the rectified waveforms, phases of relevant baseband tones are obtained. For example, in the above-mentioned case, phases of baseband tones present at 1 MHz, 2 MHz, 4 MHz, 5 MHz, and 8 MHz in the rectified $y(t)$ are extracted. Further, these extracted phases are decoded into a bit pattern according to an assigned phase margin for the symbols as

$$\text{Phase margin} = \pm\frac{\delta}{2M}. \quad (17)$$

Then the obtained bit patterns for hundreds of waveforms are compared with the transmitted bit patterns for BER calculation.

Simulated BER for multitone PSK signal transmission of 1000 bits with $N = 6$, $M = 4$, $GCD = 1$ MHz, $r = 0$ is shown in Fig. 15. As the same baseband signal which is obtained from AC-DC conversion circuitry, is being used for information as well as power, the rectifier's non-linearity introduces phase distortion in information detection.

It can be seen that BER is lowest for $\delta = [-180°, 180°]$

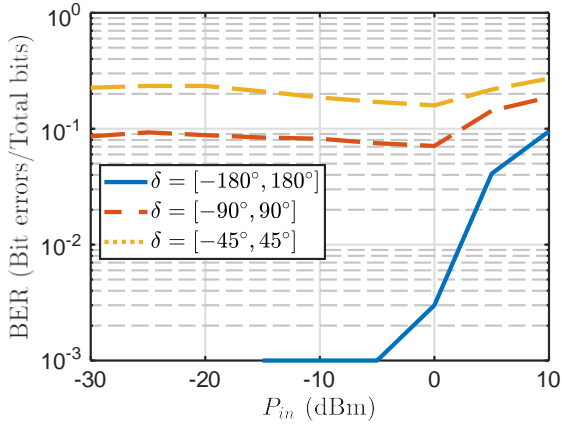

Fig. 15. Simulated BER using multitone PSK with $N = 6$, $M = 4$, $GCD = 1$ MHz, $r = 0$, and for 100 multitone streams, i.e., 1000 bits.

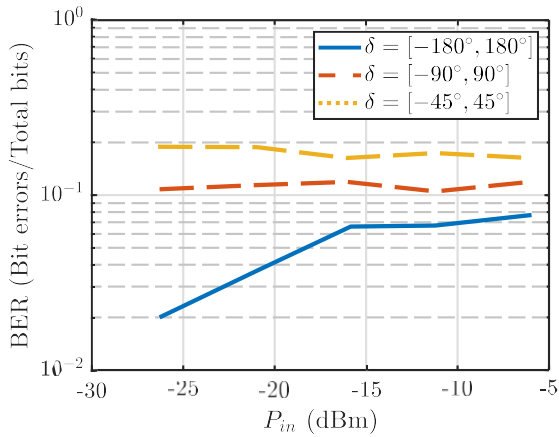

Fig. 16. Measured BER using multitone PSK with $N = 6$, $M = 4$, $GCD = 1$ MHz, $r = 0$, and for 100 multitone streams, i.e., 1000 bits.

and it is possible to have 0 BER using multitone PSK for power levels lower than −15 dBm. BER increases with the decrease in δ, and it is the highest for $\delta = [-45°, 45°]$ as allowable phase margin for correct symbol detection reduces to a narrower ±11.25° with the decreasing $\delta = [-45°, 45°]$ compared to a phase margin of ±45° for $\delta = [-180°, 180°]$. Furthermore, BER increases with increasing input power above a certain power level due to the presence of nonlinearities (diodes) in the circuit, causing AM-PM distortion and thus negatively impacting BER [29].

The simulated results are verified by the performed measurements shown in Fig 16. The increase in BER with the decreasing δ is clearly visible in the measured results. However, the measured BER is slightly higher than the simulated results. Harmonics can reflect back to the diode, and this can influence BER. The impact of these reflected harmonics and the inter-frequency interference that may lie on desired information symbols increases with increasing input power. Therefore, BER increases with increasing input power. Measured symbol constellations for $N = 6$, $M = 4$, $\delta = [-180°, 180°]$, and 1 MHz baseband tone, for different received input power levels (a) $P_{in} = -21$ dBm, (b) $P_{in} = -16$ dBm, (c) $P_{in} = -11$ dBm, and (d) $P_{in} = -6$ dBm are illustrated in Fig. 17 for 100 multitone streams, i.e., 500 symbols. The ideal normalized symbol constellation is depicted in orange and the normalized measured symbol constellations are illustrated in blue. It can be observed that phase distortion increases as input power increases. However, a low power range (below 0 dBm) is of main interest for SWIPT applications, and the effect of these factors on BER is limited at low power levels. Compensation techniques such as pre-compensation at the transmitter or post-compensation at the receiver can be introduced to reduce the AM-PM distortion.

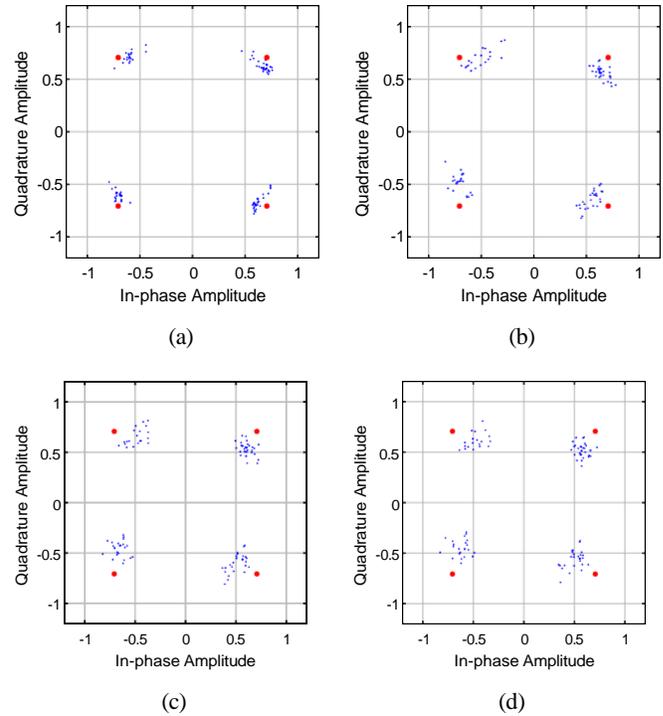

Fig. 17. Symbols constellations at baseband tone of 1 MHz for $N = 6$, $M = 4$, $\delta = [-180°, 180°]$, received input power (a) $P_{in} = -21$ dBm, (b) $P_{in} = -16$ dBm, (c) $P_{in} = -11$ dBm, and (d) $P_{in} = -6$ dBm for 100 multitone streams, i.e., 500 symbols.

Further, the effect of various $M$ over PCE and BER performances is analysed. Fig. 18 shows PCE for $N = 3$, $GCD = 1$ MHz, $r = 0$, $\delta = 360°$, and 100 multitone streams for $M = 2$, 4, and 8. It is observed that PCE does not change significantly with the increased modulation order $M$ for $N = 3$. As $M$ increases from $M = 2$ to $M = 8$, PCE increases only slightly. The reason is that now, the same symbol phase range is divided among larger number of available symbols for the case of $M = 8$ and the tone phases would be more closer to zero. Although, this effect would be more visible for a larger $N$, PCE does not vary much with $M$ compared to WIT performance.

Fig. 19 represents the simulated BER for $N = 3$, $GCD = $

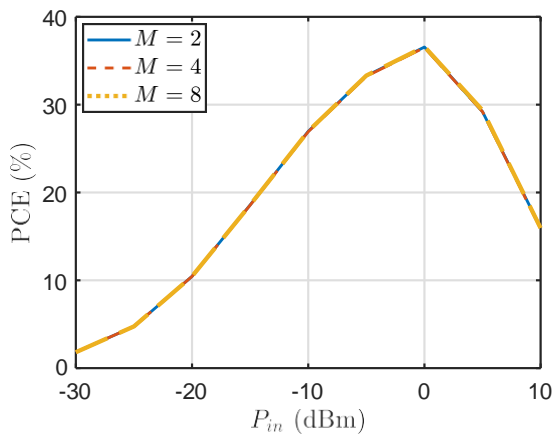

Fig. 18. Simulated PCE using multitone PSK with $N = 3$, $GCD = 1$ MHz, $\delta = 360°$, $r = 0$, and 100 multitone streams for different $M$.

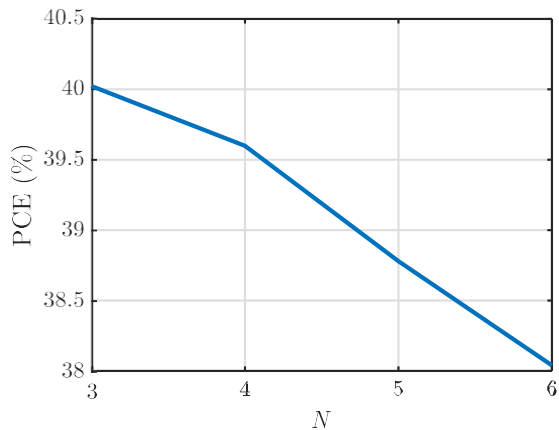

Fig. 21. Measured PCE using multitone PSK with $M = 4$, $GCD = 1$ MHz, $\delta = 360°$, $P_{in} = -6\ dBm$, $r = 0$, and 100 multitone streams for $N = 3$, $N = 4$, $N = 5$, and $N = 6$.

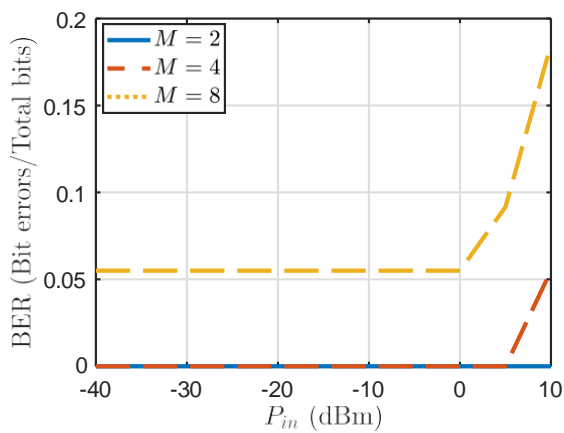

Fig. 19. Simulated BER using multitone PSK with $N = 3$, $GCD = 1$ MHz, $\delta = 360°$, $r = 0$, and 100 multitone streams for different $M$.

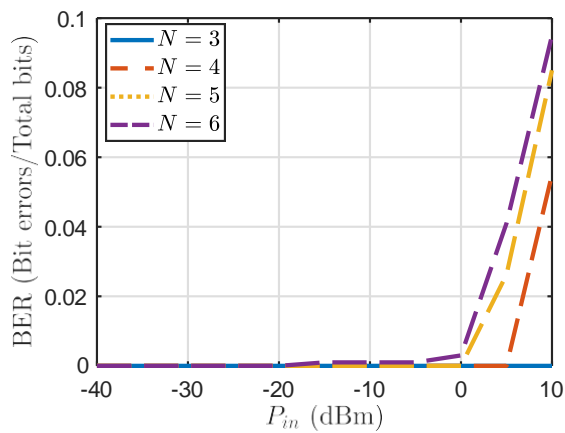

Fig. 22. Simulated BER using multitone PSK with $M = 4$, $GCD = 1$ MHz, $\delta = 360°$, $P_{in} = -6\ dBm$, $r = 0$, and 100 multitone streams for $N = 3$, $N = 4$, $N = 5$, and $N = 6$.

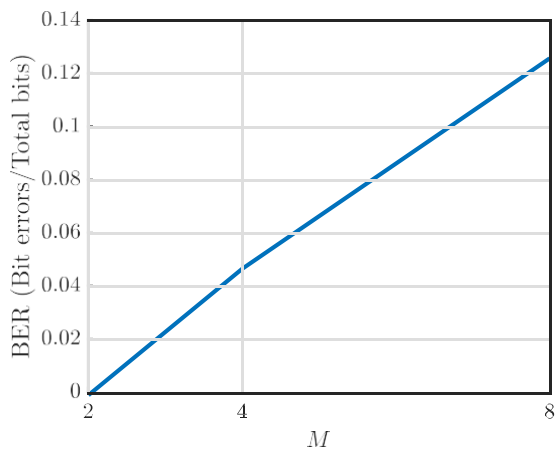

Fig. 20. Measured BER using multitone PSK with $N = 3$, $GCD = 1$ MHz, $\delta = 360°$, $P_{in} = -6\ dBm$, $r = 0$, and 100 multitone streams for $M = 2$, $M = 4$, and $M = 8$.

1 MHz, $r = 0$, $\delta = 360°$, and 100 multitone streams for each $M = 2$, 4, and 8. The corresponding measured BER performance for $N = 3$, $GCD = 1$ MHz, $\delta = 360°$, and $P_{in} = -6$ dBm at the rectifier input with 100 multitone PSK streams for various $M$ is shown in Fig. 20. It can be observed that it is possible to attain 0 BER with the lowest $M = 2$. However, BER increases from 0 to around 0.05 for $M = 4$ with the increase of just 1 bit/symbol as now, we are dividing the same $\delta$ region within the increased $M$ symbol constellation points (each with $log_2M$ bits). Therefore, phase margin per individual symbol from (17) reduces from $\pm 90°$ for $M = 2$ to $\pm 45°$ $M = 4$ for the correct detection at the output.

Table. I compares the proposed multitone PSK with the existing transmission schemes for integrated information and energy receiver architecture. Further, to analyse PCE and BER performances with varying $N$, 100 multitone PSK streams with $M = 4$, $GCD = 1$ MHz, $\delta = 360°$ for each case of $N = 3$, $N = 4$, $N = 5$, and $N = 6$ are recorded. Fig. 21 illustrates the PCE behaviour with varying $N$ for $P_{in} = -6$ dBm. It is observed that PCE reduces slightly with the increasing $N$. The reason for this is that now, more tones with misaligned phases have been included. Also, now, the multitone PSK signal BW increases with the increased $N$. For example, 6-tone multitone PSK signal has a BW of 20 MHz whereas 5-tone signal has 12 MHz. Therefore, now for $N = 6$, the required $5^{th}$ intermodulation tone at 8 MHz has lower amplitude due to the effect of LPF.

Table I: Transmission strategies for integrated information and energy receiver architecture.

| Transmission Approach | Advantages | Limitations | Power performance | Information performance |
|---|---|---|---|---|
| PAPR based [21], [22] | - High PAPR. | - Rectifier's RF BW and output filter BW increase with increasing modulation order.<br>- 1 information symbol per a single multitone signal. | - DC of 3.5 times higher compared to the single carrier input signal for 30 dB average received SNR [21].<br>- $0.5\,\mu A$ for $-10$ dB input power [22].<br>- Measurements are not performed. | - BER = $10^{-2}$ for 30 dB average received SNR [21].<br>- BER = $10^{-1}$ for $-10$ dB input power [22].<br>- Measurements are not performed. |
| Biased-ASK [24] | - Each symbol has some minimum energy. | - Single tone is used instead of a multitone signal.<br>- 1 information symbol per a single multitone signal.<br>- Modulation order limited by ripple voltage. | - 0.13 V for $M = 2$, $A_{ratio} = 0.5$, and 20 dB SNR. | - $10^{-2}$ BER for $M = 2$, $A_{ratio} = 0.5$, and SNR = 20 dB. |
| Amplitude ratio [25] | - Independent of transmission distance. | - Information detection is possible only for the multitone with a small number of tones.<br>- Modulation order limited by ripple voltage.<br>- 1 information symbol per a single multitone signal. | - 48% PCE for total input power of $-10$ dBm for tones power ration $r = 1/6$. | - WIT performance such as EVM or BER is not analyzed. |
| ADSK, ARSK [20] | - Increased operational constellation range. | - Modulation order limited by ripple voltage.<br>- 1 information symbol per a single multitone signal. | - $0.1\,\mu W$ at $-15$ dBm input power.<br>- Measurements are not performed. | - $10^{-1}$ BER for $M = 4$, and $SNR = 20$ dB.<br>- Measurements are not performed. |
| Multitone-FSK [27] | - Lessened envelope variations.<br>- Reduced impact of WIT on WPT. | - Rectifier's RF BW and output filter BW increase with increasing modulation order.<br>- 1 information symbol per a single multitone signal. | - 0.3 V for input power of $-5$ dBm. | - SER = 0.05 for $M = 4$, $r = 0.5$, and input power of 0 dBm. |
| Proposed multitone PSK in this work | - Transmission of $(N-1)$ symbols over an $N$-tone multitone signal.<br>- OFDM-type communication for WIT.<br>- Lower ripples with information embedding in tones' phases.<br>- Modulation order independent of ripple voltage.<br>- RF matched BW and output filter BW independent of modulation order.<br>- Avoiding saturation of the transmitter power amplifier for end-to-end SWIPT performance. | - RF matched BW and output filter BW increases with an increasing number of tones $N$. | - 38% PCE for $N = 3$, $M = 4$, $\delta = [-180°, 180°]$, and $P_{in} = -6$ dBm. | - BER of 0 for $M = 2$, and BER = 0.047 for $M = 4$ with $N = 3$, $\delta = [-180°, 180°]$, and $P_{in} = -6$ dBm. |

Fig. 22 shows the simulated BER behaviour with the varying $N$ and the corresponding measurement results for $P_{in} = -6$ dBm are shown in Fig. 23. BER increases with the increased $N$. This is due to the increased signal BW as now, there is more probability of having phase error with the lower amplitudes of the desired baseband tones. The effect of an increase in the above-discussed parameters such as $\delta$, $M$, and $N$ over WIT and WPT performance is listed in Table II.

Here, in this scheme, we can transmit $(N-1)$ symbols over $N$-tone signal. Therefore, the throughput can be defined by
$$T_p = \frac{(N-1).\log_2 M}{T_{PSK}}, \quad (18)$$
where $T_{PSK}$ is the time period of multitone PSK signal. From (18), it can be seen that the chosen $\delta$ does not affect the system throughput. However, an appropriate $GCD$ would be needed for the required throughput as it affects $T_{P\,SK}$. The larger is the $GCD$, the smaller would be the $T_{P\,SK}$, and in turn, the higher would be the throughput. However, BER increases with the increase in $GCD$ because now, multitone PSK signal is having larger BW, and matched circuit BW

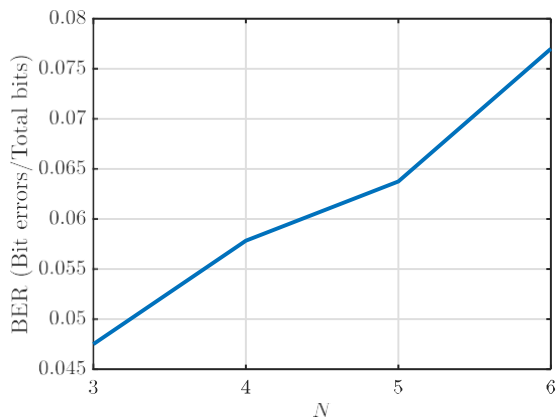

Fig. 23. Measured BER using multitone PSK with $M = 4$, $GCD = 1$ MHz, $\delta = 360°$, $P_{in} = -6$ $dBm$, $r = 0$, and 100 multitone streams for $N = 3$, $N = 4$, $N = 5$, and $N = 6$.

Table II: SWIPT performance for Multitone PSK with various parameters

| Performance parameters | PCE | BER |
|---|---|---|
| Phase range, $\delta$ | Decreases slightly (~ 3% when $\delta$ is increased from $0°$ to $[-180°, 180°]$). | Decreases ($10^{-1}$ for $\delta = [-90°, 90°]$ to $10^{-3}$ for $[-180°, 180°]$) |
| Modulation order, $M$ | does not change much | Increases (0 for $M = 2$ to 0.05 for $M = 4$) |
| Number of Tones, $N$ | Decreases slightly (~ 2% from $N = 3$ to $N = 6$) | Increases (0.047 for $N = 3$ to 0.075 for $N = 6$) |

limits the BER performance.

Multitone PSK signal is designed for information detection using only one rectifier while simultaneously utilizing the signal power. Multitone PSK signal can also be used for low-power signal communications as an information transmission scheme, but with low PCE. However, it still helps with improving end-to-end power efficiency as information reception is performed by only one rectifier. This can be a potential application in heterogeneous networks where nearby users utilize the signal for wireless power transmission and far-away users utilize it for wireless information transmission.

## VI. CONCLUSION

In this paper, a novel multitone PSK transmission scheme for integrated receiver SWIPT architecture having low power consumption has been proposed. Information is encoded in terms of phase differences of consecutive tones of the multitone signal. ($N - 1$) symbols are transmitted over a single stream of $N$-tone multitone PSK signal. The WPT and WIT performances of the proposed transmission scheme are analysed in terms of PCE and BER, respectively. The main advantage of encoding the information in tone phases instead of tone amplitudes is the lesser variation in the output ripple voltage. Therefore, increasing the modulation order mainly affects WIT performance only while keeping the WPT performance approximately the same, which is a benefit from SWIPT perspective. A suitable rectifier circuitry is fabricated and all the simulation results are verified with the measurements. The effect of various signal designing parameters such as allocated phase range, modulation order, and number of tones is analysed from both WIT and WPT perspectives for the overall SWIPT performance of a system.

In the future, more practical, wireless with time-varying scenarios together with communication performance enhancement protocols for managing distortion need to be considered. These SWIPT waveform protocols can be optimized for the users' power level or information rate requirement whichever is critical for a particular user. This would be a step towards standardizing the SWIPT waveforms for future IoT networks.